# Ransomware Classification and Detection With Machine Learning Algorithms

Mohammad Masum∗, Md Jobair Hossain Faruk†, Hossain Shahriar‡
Kai Qian§, Dan Lo§, Muhaiminul Islam Adnan¶

∗School of Data Science, Kennesaw State University, USA
†Department of Software Engineering and Game Development, Kennesaw State University, USA
‡Department of Information Technology, Kennesaw State University, USA
§Department Computer Science; Kennesaw State University, USA
¶Institute of Natural Sciences, United International University, Bangladesh.
Email: {*mmasum∗, mhossa21†*}@students.kennesaw.edu | {*hshahria‡, kqian§, dlo2§*}@kennesaw.edu
{*adnan08mt¶*}@gmail.com

*Abstract*— **Malicious attacks, malware, and ransomware families pose critical security issues to cybersecurity, and it may cause catastrophic damages to computer systems, data centers, web, and mobile applications across various industries and businesses. Traditional anti-ransomware systems struggle to fight against newly created sophisticated attacks. Therefore, state-of-the-art techniques like traditional and neural network-based architectures can be immensely utilized in the development of innovative ransomware solutions. In this paper, we present a feature selection-based framework with adopting different machine learning algorithms including neural network-based architectures to classify the security level for ransomware detection and prevention. We applied multiple machine learning algorithms: Decision Tree (DT), Random Forest (RF), Naïve Bayes (NB), Logistic Regression (LR) as well as Neural Network (NN)-based classifiers on a selected number of features for ransomware classification. We performed all the experiments on one ransomware dataset to evaluate our proposed framework. The experimental results demonstrate that RF classifiers outperform other methods in terms of accuracy, F-beta, and precision scores.**
.

## I. INTRODUCTION

Malicious applications or attacks, malware and ransomware families for instance, consistently endures to pose critical security issues to cybersecurity and it may cause catastrophic damages to computer systems, data centers, web, and mobile applications across various industries and businesses[1]–[3]. Most ransomware is designed to block and prevent targeted victims from accessing computer data by applying an indestructible encrypting methodology that can be decrypted by the attacker itself solely. Removing the ransomware leads the victim to irreversible losses, as a result, victims are forced to pay according to the attacker's demands [4]. Failure or denial to comply with the attacker's demand will lead to losing data permanently. With the help of modern technology, attackers are transforming conventional ransomware into emerging ransomware families which is more difficult in reversing a ransomware infection [5].

Ransomware is a sophisticated and variants threat affecting users worldwide that limits users from accessing their system or data, either by locking the system's screen or by encrypting and the users' files unless a ransom is paid [2]. Two primary forms of ransomware based on attack approaches include locker ransomware that denies access to the computer or device and crypto ransomware that prevents access to files or data [6]. After these attacks, it is incredibly difficult to revert without paying the extortion. Traditional ransomware detection techniques including event-based, statistical-based, and data-centric-based techniques are not adequate to combat. Therefore, implementing the highest level of optimal protection and security by adopting futuristic technology against such advanced malicious attacks should be imperative for the research community.

Novel technology, machine learning for instance in ransomware detection is a new research topic and can be immensely utilized in the development of innovative ransomware solutions [7]. Employing the application of Machine Learning (ML) methodologies enables automatic detection of malware including ransomware through their dynamic behaviors and enhances security [8]. Algorithms such as Decision Tree (DT), Random Forest (RF), Naïve Bayes (NB), Logistic Regression (LR), and Neural Network (NN)-based architectures have potential efficacy for ransomware classification and detection [9]. In this study, we conduct a comprehensive assessment and investigates the machine learning techniques for the classification of ransomware. The primary contributions of the paper as follows:

- We conduct a comprehensive investigation on classifying ransomware and propose a framework by selecting a number of features for model development process with adopting traditional ML classifiers and NN-based architectures.
- We demonstrate the generalization of the models' performance by providing robust experiments and compare it with the various methods.

The rest of the paper is organized as follows: In Section II, we discuss ML-based related work of ransomware detection. Section III explains the methods we applied in this paper. The experimental setting and results are explained in Section IV. Finally, Section V concludes the paper.

## II. RELATED WORK

Conventional detection techniques have been applied for classifying various malware including ransomware. Various ransomware can be analyzed by a well-defined behavioral structure and most of the ransomware families share common behavioral traits including payload persistence, stealth techniques, network traffic. Signature-based analysis is the most widely used traditional anti-malware system and A. M. Abiola and M. F. Marhusin [10] proposed a signature-based detection model for malware by extracting the Brontok worms and to break down the signatures, an n-gram technique was utilized. The framework enables to detection of malware and creates a credible solution that eliminates any threats. To improve the limitation, a static and dynamic-based or Behavior-based framework was introduced by [11] where analysis static-based technique analyze the application's code to determine malicious activities and dynamic-based analysis on the other hand monitoring the processes to determine the behavior of malicious intent and will be flagged as suspicious and terminated. Both static and dynamic-based analysis has limitation in terms of the inability to detect unknown malware and ineffectiveness against code obfuscation, high variant output, and targeted attacks. F. Noorbehbahani and M. Saberi [8] focused on semi-supervised learning for exploiting a number of labeled data and a lot of unlabeled data towards detecting ransomware. Different feature selection and semi-supervised classification methods were applied to the CICAndMal 2017 dataset for analyzing the ransomware and the semi-supervised classification method using the random forest as a base classifier outperforms the various semi-supervised classification techniques for ransomware detection.

In order to improve the conventional approaches, state-of-the-art machine learning concept needs to be adopted in ransomware detection and prevention. A group of researchers [12] proposed a network intrusion detection framework consisting of Argus server and client applications by introducing a novel flow-oriented method as Biflow for detecting ransomware. For the classification of the datasets, six feature selection algorithms were adopted and for achieving better accuracy and enhancing the performance of the detection module, supervised machine learning was utilized. Random Forest is one of the popular machine learning techniques that has been used for malware and ransomware detection. F. Khan et al. [13] proposed a DNAact-Ran, A Digital DNA Sequencing Engine based ransomware detection framework that focuses on sequencing design constraints and k-mer frequency vector. The framework was demonstrated on 582 DNAact-Run ransomware and 942 goodware instances to measure the performance of precision, recall, f-measure, and accuracy. S. Poudyalwe et al. [14] introduced a machine learning-based detection model to efficiently detect ransomware that adopts multi-level analysis for better interpreting the purpose of malware code segments. The model was evaluated, and the results indicate its performance in detecting ransomware between 76% to 97%. V. G. Ganta et al. [15] proposed an approach that is opposite to the traditional ransomware detection system by adopting a machine learning approach. The framework utilized different classification algorithms including ex-random forest, decision tree, logistic regression, and KNN algorithm to detect ransomware hides in executable files.

Researcher Daniele Sgandurra et al. [16], proposed a machine learning-based approach for dynamically analyzing and classifying ransomware called EldeRan that monitors the bening software activities based on possible unique signs of ransomware. Two types of ML components are used in EldeRan including feature selection and classification where the Cuckoo Sandbox environment was adopted. For retrieving, and dynamically analyzing the datasets, it utilizes the following classes: Windows API calls, Registry Key Operations, File System Operations, the set of file operations performed per File Extension, Directory Operations, Dropped Files, and Strings. The framework was demonstrated using 582 ransomware datasets from 11 different families, and 942 goodware applications that indicate the accuracy of an area under the ROC curve of 0.995. Sumith Maniath et al. [17] proposed a framework on binary sequence classification of API calls by utilizing Long-Short Term Memory (LSTM) networks to classify ransomware through its behavior. A dynamic analysis technique was adopted for extracting the API calls from the modified log in a sandbox environment. According to the evaluation, the proposed LSTM based framework achieved 96.67% accuracy in classifying the ransomware behavior automatically from a large volume of malware datasets. However, by enhancing the LSTM network, the overall accuracy can be improved further.

Such accuracy supports that ML can be a viable and effective approach to detect novel ransomware variants and families. Deep Neural Network (DNN) has the ability to solve complex detection problems and DNN can be used in detecting ransomware by constructing a novel dynamic detection method. Bayesian Hyperparameter Optimization can be adopted for Deep Neural Network-Based Network Intrusion Detection where the researchers [18] proposed a novel Bayesian optimization-based framework for the automatic optimization of hyperparameters. Recent research work is presented by Hadis Ghanei et al. [19] where a dynamic malware detection framework using Deep Neural Network (DNN) and Convolutional Neural Network (CNN) was proposed for malware detection. Long Short-Term Memory (LSTM) is used to construct the machine learning model. Between CNNs and the LSTM network, a novel approach was used for determining suspicious samples of malware. According to the evaluation report, a combination of DNN and LSTM provide effective in detecting new malware and achieved 91.63% accuracy. Deep Learning has also been used to detect malware in Android. M. Masum and H. Shahriar proposed a deep learning framework (Droid-NNet) for malware classification in Android, mechanized as a deep learner that outperforms existing cutting-edge machine learning methods. Based on the evaluation report on two Android apps datasets, Malgenome-215 and Drebin-215, the Droid-NNet indicates robust and effective malware detection in the Android platform [20].

The proposed framework was optimized with a limited number of important features and experimented with different ML classifiers including neural network-based architecture. The experimental results show the robustness and effectiveness of proposed framework.

### III. METHODOLOGY

We applied traditional ML classifiers (e.g., decision tree classifier, random forest classifier, naïve bayes classifier, and logistic regression classifier) and neural network-based architecture to detect ransomware.

Fig. 1 shows the framework of our model. The ransomware data were standardized to convert different scale variable into a similar range. Feature selection method was applied to select a number of important features from the data and consequently, feed the features into different classifiers to detect ransomware from legitimate observations. We implemented 10-fold cross validation technique to generalize the model. Finally, we reported different evaluation metrics such as accuracy, *F-beta* score, precision, recall and area under ROC curve to assess the models' performance.

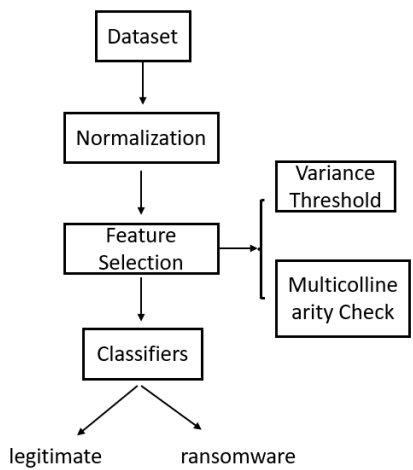

*Figure 1: Framework to detect ransomware*

### IV. EXPERIMENTS AND RESULTS

#### A. Dataset specification

The dataset contains in total 138,047 samples with 54 features and was collected from [21] where 70% are ransomware and remaining 30% are legitimate observations. Fig. 2 shows the distribution of the dataset.

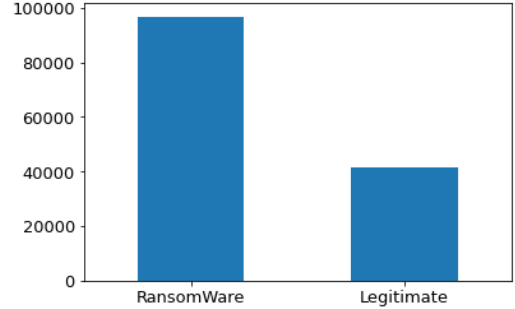

*Figure 2 : Distribution of the dataset*

#### B. Feature Selection

Z-score standardization technique was used to convert each of the variables into a similar scale by centering each of the variables at zero with a standard deviation of 1. We applied feature selection methods such as variance threshold and variance inflation factor to remove low variant and highly correlated features from the data, respectively. Removing low variant features from the dataset, a variance threshold score was set 1, since the number of features dramatically dropped from 54 to 13 when threshold was set to 1. Fig. 3 shows number of features with varying variance threshold scores.

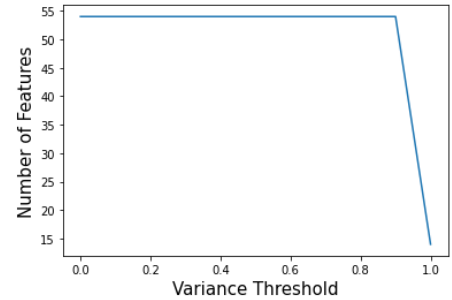

*Figure 3: Number of features with varying variance threshold*

In the second step of feature selection, we checked the multicollinearity of the high variance features using variance inflation factor (VIF). A VIF score 10 was selected to identify highly correlated features, meaning that a feature is identified if the VIF score is higher than 10. Features: SectionMeanRawSize and SectionMaxRawSize show multicollinearity by displaying 19.52 and 19.48 VIF scores, respectively. We randomly dropped one of these variables. Table 1 illustrates the 12 high variant features with associated VIF score, all of which are fall inside the VIF threshold. Finally, we feed these 12 selected variables to the classifiers to detect ransomware.

*Table 1: Selected features after applying variance threshold and VIF criterion*

| Feature | VIF |
|---|---|
| SizeOfOptionalHeader | 1.24 |
| MajorLinkerVersion | 1.15 |
| AddressOfEntryPoint | 1.04 |
| SectionAlignment | 1.03 |
| MinorOperatingSystemVersion | 4.04 |
| SizeOfHeaders | 1.0 |
| SizeOfStackReserve | 1.19 |
| LoaderFlags | 4.04 |
| SectionsMinEntropy | 1.31 |
| SectionsMaxEntropy | 1.41 |
| SectionMaxRawsize | 1.0 |
| SectionsMinVirtualsize | 1.02 |
| ResourcesMinEntropy | 1.08 |

*C. Evaluation metrics*

1. *Recall:* The number of correct positive predictions among all the positive samples. Mathematically:

$$Recall = \frac{TP}{TP + FN}$$

Where, TP is True Positive (quantity of correct positive predictions) and FN is False Negative (quantity of misclassified positive predictions)

2. *Precision:* The proportion of the correctly identified positives to all the predicted positives. Mathematically:

$$Precision = \frac{TP}{TP + FP}$$

3. $F_1$ *score :* The harmonic means of Precision and Recall. $F_1$ *score* is a better performance metric than the accuracy metric for imbalanced data.

$$F_1 = 2 \times \frac{Precision \times Recall}{Precision + Recall}$$

The F-beta score is the weighted harmonic mean of precision of recall where F-beta value at 1 means perfect score (perfect precision and recall) and 0 is worst.

$$F_\beta = (1 + \beta^2) \frac{Precision \times Recall}{(\beta^2 \times Precision) + Recall}$$

When $\beta = 1$, F-beta is $F_1$ *score*. The $\beta$ parameter determines the weight of precision and recall. $\beta < 1$ can be picked, if we want to give more weight to precision, while $\beta > 1$ values give more weight to recall.

*D. Experimental setting*

We evaluated our model performance by comparing it with the performance of LR, NB, RF, and DT methods. Both datasets were randomly split into training and test data while maintaining the class ratio between legitimate and ransomware samples. Trained data was used to train each of the models we experimented with while test data was used for evaluating the performance of the models. To verify the consistency of the model, we experimented with each of the models with 10-fold cross-validation.

The RF, LR, NB, DT and NN-based classifiers were applied to both datasets for comparing results with our framework. The algorithms were implemented using Python scikit-learn library with available hyperparameter options.

The neural network-based architecture consists of 4 layers including one input layer, two hidden layers and one output layer. We used 'ReLu' activation function in the hidden layers and 'sigmoid' function in the output layer, as this is a binary classification problem. 'Adam' and 'binary cross-entropy' were used for optimizer and loss function respectively. We implemented an early stopping method to stop training once the model performance stops improving on the test data. We selected validation loss to be monitored for early stopping and set minimum delta to $1e - 3$ (checks minimum change in the monitored quantity to qualify as an improvement) and patience to 5 (checks number of epochs that produced the monitored quantity with no improvement after which training will be stopped). The initial learning rate was set to 0.01

*E. Results*

We applied DT, RF, NB, LR, and NN classifiers to classify between legitimate and ransomware samples. Table **3** demonstrates the results of the models in terms of accuracy, F-beta score, recall and precision. Random Forest classifier outperforms other models by achieving highest accuracy, F-beta score and precision. NB classifier achieves highest recall, though it provides poor performance in terms of other performance metrics. Both DT and NN classifiers shows reasonable performance compared to RF. However, LR fails to achieve rewarding F-beta score and recall score compared to other methods, though the accuracy score is reasonable compared to DT, RF, and NN classifiers. Fig 4-8 illustrates the ROC curve for each of the classifiers with containing 10-fold curves and mean curve. The RF, LR, and NN achieved identical maximum mean Area Under Curve (AUC) score of 0.99 while the lowest was achieved by NB (mean AUC: 0.73)

*Table 2: Experimental results analysis of different classifiers*

| Classifiers | Accuracy | F-beta | Recall | Precision |
|---|---|---|---|---|
| DT | 0.98±0.01 | 0.94±0.05 | 0.94±0.05 | 0.98±0.00 |
| RF | **0.99±0.01** | **0.97±0.03** | 0.97±0.03 | **0.99±0.00** |
| NB | 0.35±0.03 | 0.97±0.03 | **0.99±0.00** | 0.31±0.01 |
| LR | 0.96±0.02 | 0.89±0.07 | 0.89±0.07 | 0.96±0.00 |
| NN | 0.97±0.01 | 0.95±0.05 | 0.95±0.05 | 0.97±0.00 |

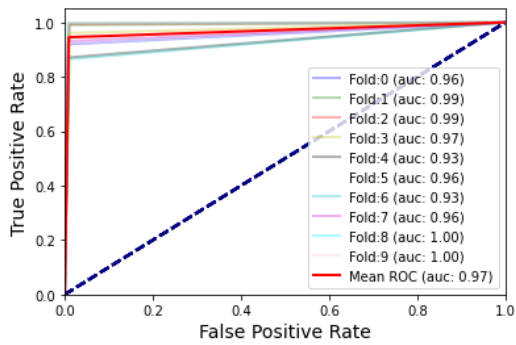

*Figure 4: ROC curve for Decision Tree classifier*

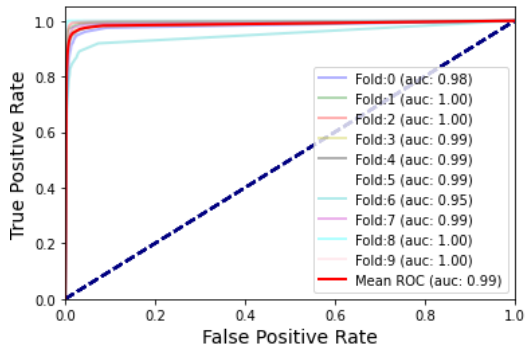

*Figure 5: ROC curve for Random Forest classifier*

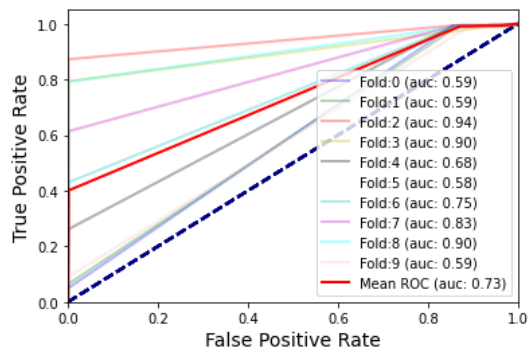

*Figure 6: ROC curve for Naïve Bayes classifier*

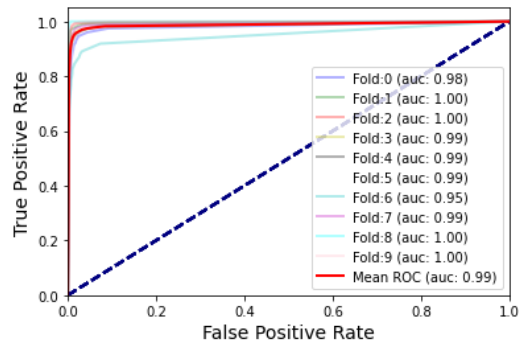

*Figure 7: ROC curve for Logistic Regression classifier*

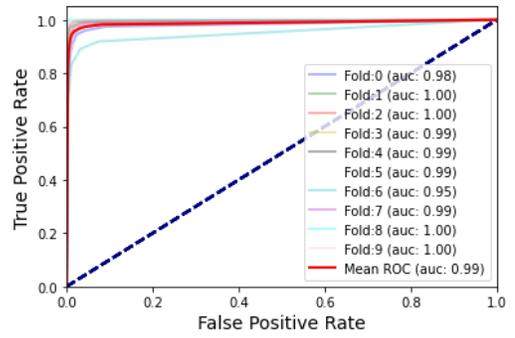

*Figure 8: ROC curve for Neural Network classifier*

## V. Conclusion

Malware including ransomware is increasingly posing a serious security threat to financial institutions, businesses, and individuals. It is essential to develop an automatic system to effectively classify and detect ransomware and reduce the risk of malicious activities. In this paper, we presented a feature selection-based novel framework, adopted different machine learning algorithms including neural network-based classifiers for effective ransomware classification and detection. We applied the framework with all the experiments on a ransomware dataset and evaluated the models' performance by a robust comparative analysis among DT, RF, NB, LR, and NN classifiers. The experimental results demonstrate that the Random Forest classifier outperformed other classifiers by achieving the highest accuracy, F-beta, and precision scores with reasonable consistency in the 10-fold cross-validation results.

## Acknowledgement

The work is partially supported by the U.S. National Science Foundation Awards #2100134, #2100115, #1723578, #1723586. Any opinions, findings, and conclusions or recommendations expressed in this material are those of the authors and do not necessarily reflect the views of the National Science Foundation.